\documentclass[12pt]{article}
\usepackage{graphicx} 

\usepackage{setspace}
\usepackage{bbm}
\usepackage{rotating}
\usepackage{amsmath}
\usepackage{multirow}
\usepackage{latexsym,amssymb,amsmath}
\usepackage{CJK}
\usepackage{picinpar,graphicx}
\usepackage{float}
\usepackage{listings}
\usepackage{color}
\usepackage{bm}
\usepackage{mathrsfs}
\usepackage{graphicx}
\usepackage{subfigure}
\usepackage{epstopdf}
\usepackage{multirow}
\usepackage{authblk}
\usepackage{threeparttable}
\usepackage{booktabs} 
\usepackage[authoryear]{natbib}

\newtheorem{definition}{Definition}
\newtheorem{theorem}{Theorem}

\begin{document}

\title{Subgroup analysis in randomized controlled trials with binary outcomes: dilution and logic-respecting properties}

\author[1]{Long-Hao Xu}
\author[2]{Yang Han\footnote{Corresponding author: {\sf{e-mail: yang.han@manchester.ac.uk}}}}
\author[1]{Tim Friede}

\affil[1]{Department of Medical Statistics, University Medical Center G\"{o}ttingen, G\"{o}ttingen, Germany}
\affil[2]{Department of Mathematics, The University of Manchester, Manchester, UK}

\date{\today} 


\maketitle

\begin{abstract}

Subgroup analysis is routinely used in randomized controlled trials to examine whether treatment effects are homogeneous across patient subgroups or differ because of treatment-effect heterogeneity. In this paper, we investigate the properties of the odds ratio and the relative response in subgroup analyses with binary outcomes, extending previous work with new theoretical insights and methodological developments. We establish several new theorems that characterize how the odds ratio for the overall population changes in both magnitude and direction when two subgroups are combined. These results further confirm that the odds ratio is inappropriate as an efficacy measure in this subgroup setting, whereas the relative response is appropriate. We also present the formal relationship between the odds ratio and the relative response, and clarify their differences in terms of the logic-respecting property, that is, whether the overall efficacy lies between the subgroup efficacies, and the dilution property, that is, whether mixing subgroups moves the overall odds ratio toward 1. Although the odds ratio is generally not logic-respecting, it may behave approximately like a logic-respecting efficacy measure under certain conditions. To illustrate our findings, we present an illustrative example based on clinical trial data and discuss its implications for subgroup analysis in randomized controlled trials.

\end{abstract}


\section{Introduction}\label{sec-1}

In recent years, precision medicine has received increasing attention in the field of drug development (cf. \citealp{W2016, DPG2018}).
The goal of precision medicine is to accurately identify patient characteristics and to assign treatments that are most effective for individual patients.
In drug development, randomized controlled trials (RCTs) remain the gold standard for evaluating whether a new drug or therapy is effective compared to a control.
The relative difference in outcomes between the treatment and control groups is referred to as the efficacy of the new intervention.

An important challenge in precision medicine is determining whether a treatment should be administered to all patients or only to specific subgroups.
A comprehensive systematic review on this topic can be found in \cite{O2016}.
The efficacy of a new drug or therapy may vary considerably across different patient subgroups.
In some cases, a treatment may be beneficial for one subgroup while being ineffective or even harmful for others.
For example, in the study of advanced non-small-cell lung cancer, patients with squamous cell histology were found to receive less benefit from cisplatin plus pemetrexed treatment (cf. \citealp{S2008}).

As illustrated in the example above, the overall patient population can be divided into subgroups based on specific indicators or characteristics, known as biomarkers.
Once subgroups are defined, efficacy can be analyzed both within subgroups and across the entire population.
However, certain efficacy measures may yield paradoxical results, thereby complicating the decision-making process (cf. \citealp[Section 2]{Liu2020}).
Consider the following scenario: the efficacy of a new therapy is considered positive within two distinct subgroups.
However, when these subgroups are combined into a single overall population, the same therapy is evaluated as ineffective.
Given that the overall population is composed of these two subgroups, it is reasonable to expect that if the therapy demonstrates efficacy in both, it should likewise be effective in the combined population.
This discrepancy introduces a logical paradox, which challenges the validity of the efficacy measure.
Consequently, such inconsistency limits our ability to reach reliable conclusions about the true effectiveness of the therapy.

To address such paradoxes, the concept of a logic-respecting efficacy measure was first introduced in the context of precision medicine by \cite{DLH2016}.
Intuitively, it is reasonable to expect that the overall efficacy should lie between the minimum and maximum subgroup efficacies.
Although this idea was formalized more recently, its roots can be traced back to earlier discussions on contingency tables in \cite{GM1987}.
It is important to emphasize that the purpose of a logic-respecting efficacy measure is to prevent such paradoxes and ensure consistency in subgroup analysis.

Researchers have explored logic-respecting efficacy measures across various types of outcomes, including continuous outcomes \citep{HTLH2021, Liu2016}, binary outcomes \citep{LXDH2019, Liu2020}, and time-to-event outcomes \citep{DLH2016, WHCCD2020, Liu2020}.
A comprehensive review of this literature is provided in \cite{Liu2020} and \cite{CDDH2021}.
In general, neither the odds ratio nor the hazard ratio satisfies the logic-respecting property (cf. \cite{Liu2020}).
To enhance logic-respecting properties, \cite{PX2020} proposed alternative definitions for the odds ratio and hazard ratio, which were further extended in \cite{LWTH2021}.
Moreover, \cite{TKH2013} provided guidance on study design in the presence of patient subgroups.
\cite{DS2022} provided additional insights into logic-respecting efficacy measures from a causal perspective.
More recently, \cite{Colnet2024} discussed different efficacy measures and compared their properties in the framework of causal inference.

Although \cite{Liu2020} and \cite{LXDH2019} investigate important properties of the odds ratio and relative response for binary outcomes in the RCTs setting, several key aspects remain unaddressed.
In particular, \cite{Liu2020} leaves ambiguity regarding the dilution property of the odds ratio when two subgroups are combined.
Moreover, neither study clarifies the precise relationship between the odds ratio and the relative response.
In this paper, we focus on efficacy measures for binary outcomes and systematically compare the odds ratio and the relative response with respect to the logic-respecting property and the dilution property.
Motivated by \cite{Liu2020}, we establish a more general result stated as Theorem \ref{theorem-g-unchange} which characterizes the direction of dilution and the ordering of the overall odds ratio under several configurations, when two subgroups are combined.
This can be viewed as a natural generalization of the original theorem in \cite{Liu2020}.
We further investigate the possibility of multiple solutions to subgroup arising from the dilution effect and present new insights in Theorem \ref{theorem-root}.
Finally, we examine the logic-respecting property of the relative response and formalize its relationship with the odds ratio.
In Theorem \ref{theorem-5}, we show that, under a specific condition, the odds ratio approximates a logic-respecting measure of efficacy.
These results further support the view that the odds ratio may be inappropriate as an efficacy measure in this setting, while the relative response appears to be more suitable, as previously noted in \cite{LXDH2019} and \cite{Liu2020}.

This paper is organized as follows.
Section \ref{sec-2} introduces the notation and preliminaries related to RCTs and subgroup analysis.
Section \ref{sec-3} proposes the new properties of the odds ratio and relative response in subgroup analysis, establishes the relationship between them, and investigates the approximation technique to the odds ratio in subgroup analysis.
Section 4 presents an illustrative example based on clinical trial data. 
Finally, Section \ref{sec-5} concludes the paper and discusses the outlook for subgroup analysis in RCTs based on our findings.
All proofs are given in the Supplementary Material.

\section{Notation and preliminaries}\label{sec-2}

In this section, we begin by introducing the fundamental concepts of subgroup analysis and provide a formal definition of a logic-respecting efficacy measure.
We then present the notation for binary outcomes in RCTs, which will serve as the foundation for the subsequent analysis.
At the end of this section, we present the definitions of the odds ratio and the relative response, which are the main focus of this paper.
As will be shown, the odds ratio does not satisfy the logic-respecting property, whereas the relative response does.

In RCTs, patients are randomly assigned to one of two groups: the treatment group, which receives the new therapy or drug (denoted by $Rx$), and the control group, which receives a placebo or no intervention (denoted by $C$).
The relative treatment effect between $Rx$ and $C$ is used to assess the efficacy of the new intervention.
As discussed in Section \ref{sec-1}, the primary goal of precision medicine is to identify those patients who are more likely to benefit from the new treatment.
To achieve this, biomarkers are often used to stratify the patient population into subgroups.
Biomarkers can take many forms, including genetic variations, concentrations of specific biological substances, or other measurable biological characteristics.
For example, a single nucleotide polymorphism may be used to divide patients into three genotype-based subgroups: $AA$, $Aa$, and $aa$.

In the general setting, the overall patient population can be partitioned into two mutually exclusive subgroups: $g^+$ subgroup and its complement $g^-$ subgroup, defined by a specific biomarker.
Since the total population $\{g^+,g^-\}$ is a mixture of these two subgroups, it is natural to expect that the overall treatment efficacy in the overall population lies between the efficacies observed in the two subgroups.
This intuitive property forms the basis of the definition introduced below.
For clarity, we adopt the notation established in \citet{Liu2020} and \citet{LXDH2019}, where the true treatment efficacies in $g^+$ subgroup, the $g^-$ subgroup, and the overall population $\{g^+,g^-\}$ are denoted by $\mu_{g^+}$, $\mu_{g^-}$, and $\mu_{\{g^+,g^-\}}$, respectively.
Without loss of generality, we assume $\mu_{g^-}\leq\mu_{g^+}$.

\begin{definition}\label{logic-respecting}
An efficacy measure is \textbf{logic-respecting} if
\begin{align*}
\mu_{\{g^+,g^-\}}\in[\mu_{g^-},\mu_{g^+}].
\end{align*}
\end{definition}

The concept of logic-respecting efficacy measure was first introduced in the context of drug development by \cite{DLH2016}.
For the case $\mu_{g^-}<\mu_{g^+}$, the efficacy in the overall population can be expressed as:
\begin{align*}
\mu_{\{g^+,g^-\}}=\frac{\mu_{\{g^+,g^-\}}-\mu_{g^-}}{\mu_{g^+}-\mu_{g^-}}\mu_{g^+}+\frac{\mu_{g^+}-\mu_{\{g^+,g^-\}}}{\mu_{g^+}-\mu_{g^-}}\mu_{g^-}.
\end{align*}
This representation shows that if an efficacy measure is logic-respecting, then $\mu_{\{g^+,g^-\}}$ can be interpreted as a convex combination (i.e., a weighted average with non-negative weights) of the subgroup efficacies $\mu_{g^+}$ and $\mu_{g^-}$.
Specifically, the weight assigned to $\mu_{g^+}$ is $w=\frac{\mu_{\{g^+,g^-\}}-\mu_{g^-}}{\mu_{g^+}-\mu_{g^-}}\geq0$, and the corresponding weight for $\mu_{g^-}$ is $1-w\geq0$.
In the special case $\mu_{g^-}=\mu_{g^+}$, the logic-respecting property implies that the overall efficacy must also equal the common subgroup efficacy, i.e., $\mu_{g^-}=\mu_{g^+}=\mu_{\{g^+,g^-\}}$.
It is important to emphasize that the logic-respecting property is a property of the efficacy measure itself, not of the underlying statistical model (cf. \citealp{CDDH2021}).
This distinction underscores the need to carefully evaluate the choice of efficacy measure in subgroup analyses to ensure consistent and interpretable results.

In this paper, we primarily focus on binary outcomes in the presence of two mutually exclusive subgroups, denoted as the $g^+$ and $g^-$ subgroups.
An important assumption for analyzing binary outcomes in RCTs is that the proportion of individuals in the $g^+$ subgroup within the overall population is independent of treatment assignment.
That is, the probability of an individual belonging to the $g^+$ subgroup is the same across both the treatment ($Rx$) and control ($C$) groups: $\mathbb{P}(g^+|C)=\mathbb{P}(g^+|Rx)=\gamma^+$, where $\gamma^+$ denotes the prevalence of the $g^+$ subgroup in the population.

We represent the binary outcomes using joint probabilities as shown in Table \ref{T-joint}.
Here, $R$ denotes response and $NR$ denotes non-response.
The notation used in this table is straightforward and intuitively interpretable.
For example, $p_{g^+}^{Rx}(R)$ denotes the probability of response among patients in the $g^+$ subgroup who receive the treatment $Rx$.
Similarly, $p_{g^-}^{C}(R)$ denotes the probability of response among patients in the $g^-$ subgroup who receive the control treatment $C$.
It is important to note that the joint probabilities for the overall population $\{g^+,g^-\}$ can be derived by summing the corresponding probabilities from each subgroup.
For instance, the overall response probability under the control treatment is given by $p^{C}(R)=p^{C}_{g^+}(R)+p^{C}_{g^-}(R)$, which indicates that the total probability of response in the control group is the sum of the response probabilities from the $g^+$ and $g^-$ subgroups within the control arm.

\begin{table}[htbp]
\centering
\footnotesize
\caption{Joint probabilities for the \(g^+\), \(g^-\), and \(\{g^+,g^-\}\) populations}
\label{T-joint}
\setlength{\tabcolsep}{4pt}
\renewcommand{\arraystretch}{1.15}
\begin{tabular}{lccc@{\hspace{10pt}}ccc@{\hspace{10pt}}ccc}
\toprule
& \multicolumn{3}{c}{\(g^+\) subgroup}
& \multicolumn{3}{c}{\(g^-\) subgroup}
& \multicolumn{3}{c}{\(\{g^+,g^-\}\) population} \\
\cmidrule(r){2-4} \cmidrule(r){5-7} \cmidrule(l){8-10}
& \(R\) & \(NR\) & Total
& \(R\) & \(NR\) & Total
& \(R\) & \(NR\) & Total \\
\midrule
\(Rx\) 
& \(p_{g^+}^{Rx}(R)\) & \(p_{g^+}^{Rx}(NR)\) & \(p_{g^+}^{Rx}\)
& \(p_{g^-}^{Rx}(R)\) & \(p_{g^-}^{Rx}(NR)\) & \(p_{g^-}^{Rx}\)
& \(p^{Rx}(R)\) & \(p^{Rx}(NR)\) & \(p^{Rx}\) \\

\(C\)
& \(p_{g^+}^{C}(R)\) & \(p_{g^+}^{C}(NR)\) & \(p_{g^+}^{C}\)
& \(p_{g^-}^{C}(R)\) & \(p_{g^-}^{C}(NR)\) & \(p_{g^-}^{C}\)
& \(p^{C}(R)\) & \(p^{C}(NR)\) & \(p^{C}\) \\

Total
& \(p_{g^+}(R)\) & \(p_{g^+}(NR)\) & \(p_{g^+}\)
& \(p_{g^-}(R)\) & \(p_{g^-}(NR)\) & \(p_{g^-}\)
& \(p(R)\) & \(p(NR)\) & \(1\) \\
\bottomrule
\end{tabular}
\end{table}

\begin{table}[htbp]
\centering
\footnotesize
\caption{Conditional probabilities for the \(g^+\), \(g^-\), and \(\{g^+,g^-\}\) populations}
\label{T-condi}
\setlength{\tabcolsep}{5pt}
\renewcommand{\arraystretch}{1.1}
\begin{tabular}{lcc@{\hspace{10pt}}cc@{\hspace{10pt}}cc}
\toprule
& \multicolumn{2}{c}{\(g^+\) subgroup}
& \multicolumn{2}{c}{\(g^-\) subgroup}
& \multicolumn{2}{c}{\(\{g^+,g^-\}\) population} \\
\cmidrule(lr){2-3} \cmidrule(lr){4-5} \cmidrule(lr){6-7}
& \(R\) & \(NR\) & \(R\) & \(NR\) & \(R\) & \(NR\) \\
\midrule
\(Rx\) & \(p_3\) & \(1-p_3\) & \(p_4\) & \(1-p_4\) & \(p_6\) & \(1-p_6\) \\
\(C\)  & \(p_1\) & \(1-p_1\) & \(p_2\) & \(1-p_2\) & \(p_5\) & \(1-p_5\) \\
\bottomrule
\end{tabular}
\end{table}

In addition to joint probabilities, binary outcomes in RCTs can also be represented using conditional probabilities. Following the notation introduced in \cite{LXDH2019}, we define:
\begin{align*}
p_1&=\mathbb{P}(Y=1|C,g^+),\ p_2=\mathbb{P}(Y=1|C,g^-),\\
p_3&=\mathbb{P}(Y=1|Rx,g^+),\ p_4=\mathbb{P}(Y=1|Rx,g^-),
\end{align*}
where $Y=1$ indicates that the subject responds.
Under the assumption that the subgroup distribution is independent of treatment assignment, i.e., $\mathbb{P}(g^+|C)=\mathbb{P}(g^+|Rx)=\gamma^+$, we apply the law of total probability  to derive the overall response probabilities:
\begin{align*}
p_5&=\mathbb{P}(Y=1|C)\\
&=\mathbb{P}(Y=1|C,g^+)\mathbb{P}(g^+|C)+\mathbb{P}(Y=1|C,g^-)\mathbb{P}(g^-|C)\\
&=\gamma^+p_1+(1-\gamma^+)p_2,\\
p_6&=\mathbb{P}(Y=1|Rx)\\
&=\mathbb{P}(Y=1|Rx,g^+)\mathbb{P}(g^+|Rx)+\mathbb{P}(Y=1|Rx,g^-)\mathbb{P}(g^-|Rx)\\
&=\gamma^+p_3+(1-\gamma^+)p_4.
\end{align*}
Here, $p_5$ and $p_6$ can be interpreted as the weighted average (or natural mixture) of the response probabilities within the control and treatment arms, respectively.
The conditional probability representation of binary outcomes based on this notation is summarized in Table \ref{T-condi}.

As shown in Tables \ref{T-joint} and \ref{T-condi}, these two forms of representation play distinct roles depending on the analytical objective.
Specifically, the joint probability representation (Table \ref{T-joint}) is more suitable for assessing whether a given efficacy measure satisfies the logic-respecting property, while the conditional probability representation (Table \ref{T-condi}) is more convenient for analytical derivations.

The odds ratio ($OR$) is defined as the ratio of the odds of response in $Rx$ treatment and the odds of response in $C$ treatment.
Denote the odds ratio of $g^+$ subgroup, the odds ratio of $g^-$ subgroup, and the odds ratio of the total population $\{g^+,g^-\}$ as $OR_{g^+}$, $OR_{g^-}$, and $OR_{\{g^+,g^-\}}$, respectively. In terms of the conditional probability shown in Table \ref{T-condi}, we have
\begin{align}\label{OR-condi}
OR_{g^+}=\frac{f(p_3)}{f(p_1)},\quad OR_{g^-}=\frac{f(p_4)}{f(p_2)},\quad OR_{\{g^+,g^-\}}=\frac{f(p_6)}{f(p_5)},
\end{align}
where $f(x)=\frac{x}{1-x},\ x\in(0,1)$.
If we use the joint probability shown in Table \ref{T-joint}, we have
\begin{align*}
OR_{g^+}&=\frac{p_{g^+}^{Rx}(R)\times p_{g^+}^{C}(NR)}{p_{g^+}^{C}(R)\times p_{g^+}^{Rx}(NR)},\ OR_{g^-}=\frac{p_{g^-}^{Rx}(R)\times p_{g^-}^{C}(NR)}{p_{g^-}^{C}(R)\times p_{g^-}^{Rx}(NR)},\\
OR_{\{g^+,g^-\}}&=\frac{p^{Rx}(R)\times p^{C}(NR)}{p^{C}(R)\times p^{Rx}(NR)}.
\end{align*}
Note that the odds ratio is not logic-respecting (cf. \cite{Liu2020}).

The relative response ($RR$) is defined as the ratio of the probability of response in the $Rx$ group to the probability of response in the $C$ group.
Denote $RR_{g^+}$, $RR_{g^-}$, and $RR_{\{g^+,g^-\}}$ as the relative response in $g^+$ subgroup, $g^-$ subgroup, and the total population ${\{g^+,g^-\}}$, respectively. 
In terms of conditional probability shown in Table \ref{T-condi}, we have
\begin{align*}
RR_{g^+}=\frac{p_3}{p_1},\quad RR_{g^-}=\frac{p_4}{p_2},\quad RR_{\{g^+,g^-\}}=\frac{p_6}{p_5}.
\end{align*}

To express the relationship between the relative response in subgroups and the relative response in the total population, we need the help of the joint probabilistic form shown in Table \ref{T-joint}.
Using the joint probability, the relative response for $g^+$ subgroup, $g^-$ subgroup, and the total population can be calculated as
\begin{align*}
RR_{g^+}=\frac{p_{g^+}^{Rx}(R)/p_{g^+}^{Rx}}{p_{g^+}^{C}(R)/p_{g^+}^C}, \ 
RR_{g^-}=\frac{p_{g^-}^{Rx}(R)/p_{g^-}^{Rx}}{p_{g^-}^{C}(R)/p_{g^-}^C}, \ 
RR_{\{g^+,g^-\}}=\frac{p^{Rx}(R)/p^{Rx}}{p^{C}(R)/p^C}. 
\end{align*}
From \cite{LXDH2019}, we have
\begin{align}
\frac{p_{g^+}^{C}(R)}{p^{C}(R)}\times RR_{g^+}+\frac{p_{g^-}^{C}(R)}{p^{C}(R)}\times RR_{g^-}=RR_{\{g^+,g^-\}},
\end{align}
which indicates that $RR_{\{g^+,g^-\}}$ is a weighted average of $RR_{g^+}$ and $RR_{g^-}$. In other words, the relative response is logic-respecting.


\section{Main results}\label{sec-3} 

This section presents a detailed investigation of the properties of the odds ratio and relative response in subgroup analysis, including the dilution properties and approximate logic-respecting properties. Furthermore, we examine the relationship between these two efficacy measures.

\subsection{The dilution properties of the odds ratio in subgroup analysis}

\begin{table}[htbp]
\centering
\footnotesize
\caption{Conditional probabilities and odds ratios under \((p_1^{(1)},p_2,p_3^{(1)},p_4)\)}
\label{T-condi-1}
\setlength{\tabcolsep}{5pt}
\renewcommand{\arraystretch}{1.12}
\begin{tabular}{lcc@{\hspace{10pt}}cc@{\hspace{10pt}}cc@{\hspace{12pt}}}
\toprule
& \multicolumn{2}{c}{\(g^+\) subgroup}
& \multicolumn{2}{c}{\(g^-\) subgroup}
& \multicolumn{2}{c}{\(\{g^+,g^-\}\) population} \\
\cmidrule(lr){2-3} \cmidrule(lr){4-5} \cmidrule(lr){6-7}
& \(R\) & \(NR\) & \(R\) & \(NR\) & \(R\) & \(NR\)  \\
\midrule
\(Rx\) 
& \(p_3^{(1)}\) & \(1-p_3^{(1)}\)
& \(p_4\) & \(1-p_4\)
& \(p_6^{(1)}\) & \(1-p_6^{(1)}\) \\

\(C\)  
& \(p_1^{(1)}\) & \(1-p_1^{(1)}\)
& \(p_2\) & \(1-p_2\)
& \(p_5^{(1)}\) & \(1-p_5^{(1)}\) \\
\midrule
Odds ratio
& \multicolumn{2}{c}{\(OR_{g^+}^{(1)}\)}
& \multicolumn{2}{c}{\(OR_{g^-}\)}
& \multicolumn{2}{c}{\(OR_{\{g^+,g^-\}}^{(1)}\)} \\
\bottomrule
\end{tabular}
\end{table}

\begin{table}[htbp]
\centering
\footnotesize
\caption{Conditional probabilities and odds ratios under \((p_1^{(2)},p_2,p_3^{(2)},p_4)\)}
\label{T-condi-2}
\setlength{\tabcolsep}{5pt}
\renewcommand{\arraystretch}{1.12}
\begin{tabular}{lcc@{\hspace{10pt}}cc@{\hspace{10pt}}cc}
\toprule
& \multicolumn{2}{c}{\(g^+\) subgroup}
& \multicolumn{2}{c}{\(g^-\) subgroup}
& \multicolumn{2}{c}{\(\{g^+,g^-\}\) population} \\
\cmidrule(lr){2-3} \cmidrule(lr){4-5} \cmidrule(lr){6-7}
& \(R\) & \(NR\) & \(R\) & \(NR\) & \(R\) & \(NR\) \\
\midrule
\(Rx\) 
& \(p_3^{(2)}\) & \(1-p_3^{(2)}\)
& \(p_4\) & \(1-p_4\)
& \(p_6^{(2)}\) & \(1-p_6^{(2)}\) \\

\(C\)  
& \(p_1^{(2)}\) & \(1-p_1^{(2)}\)
& \(p_2\) & \(1-p_2\)
& \(p_5^{(2)}\) & \(1-p_5^{(2)}\) \\
\midrule
Odds ratio
& \multicolumn{2}{c}{\(OR_{g^+}^{(2)}\)}
& \multicolumn{2}{c}{\(OR_{g^-}\)}
& \multicolumn{2}{c}{\(OR_{\{g^+,g^-\}}^{(2)}\)} \\
\bottomrule
\end{tabular}
\end{table}

Theorem 1 in \cite{Liu2020} shows that mixing dilutes the odds ratio; specifically, when two subgroups share the same odds ratio, the odds ratio of their mixture will be closer to 1.
For clarity, we restate this result as follows: if $OR_{g^+} = OR_{g^-} = \lambda \neq 1$, then the overall odds ratio $OR_{\{g^+, g^-\}}$ is closer to 1.
Although Theorem 1 in \cite{Liu2020} establishes that the odds ratio is diluted after mixing, it does not provide a precise characterization of how this dilution occurs or indicate its direction.
This motivates us to further investigate the mechanism and implications of the dilution effect.
We use superscripts to distinguish the two situations. For example, $p_1^{(1)}, p_3^{(1)}, p_5^{(1)}, p_6^{(1)}$ denote the corresponding parameters of Situation 1 in Table \ref{T-condi-1}, whereas $p_1^{(2)}, p_3^{(2)}, p_5^{(2)}, p_6^{(2)}$ denote the corresponding parameters of Situation 2 in Table \ref{T-condi-2}.
As illustrated in Tables \ref{T-condi-1} and \ref{T-condi-2}, we begin by fixing the $g^{-}$ subgroup and varying the $g^{+}$ subgroup.
Specifically, under the condition $OR_{g^+}^{(1)}=OR_{g^+}^{(2)}=OR_{g^-}$, we examine how $OR_{\{g^+,g^-\}}$ changes.
We then present the following theorem, which can be regarded as a generalization of Theorem 1 in \cite{Liu2020}.
The detailed proof is provided in Appendix~S3 of the Supplementary Material.

\begin{theorem}\label{theorem-g-unchange}
Suppose that the $g^{-}$ subgroup is fixed and the $g^+$ subgroup varies such that $OR_{g^+}^{(1)}=OR_{g^+}^{(2)}=OR_{g^-}=\lambda$. We have the following results:
\begin{enumerate}
  \renewcommand\theenumi{\roman{enumi})}
  \renewcommand\labelenumi{\theenumi}
  \item If $p_4>p_2$ and $p_2<p_1^{(1)}<p_1^{(2)}$, then $\lambda>OR_{\{g^+,g^-\}}^{(1)}>OR_{\{g^+,g^-\}}^{(2)}>1$.

  \item If $p_4>p_2$ and $p_1^{(1)}<p_1^{(2)}<p_2$, then $1<OR_{\{g^+,g^-\}}^{(1)}<OR_{\{g^+,g^-\}}^{(2)}<\lambda$.

  \item If $p_4<p_2$ and $p_2<p_1^{(1)}<p_1^{(2)}$, then $\lambda<OR_{\{g^+,g^-\}}^{(1)}<OR_{\{g^+,g^-\}}^{(2)}<1$.

  \item If $p_4<p_2$ and $p_1^{(1)}<p_1^{(2)}<p_2$, then $1>OR_{\{g^+,g^-\}}^{(1)}>OR_{\{g^+,g^-\}}^{(2)}>\lambda$.
\end{enumerate}
\end{theorem}

Theorem \ref{theorem-g-unchange} provides valuable insights into how the overall odds ratio changes when subgroups with identical odds ratios are combined.
This result not only confirms the dilution phenomenon mathematically but also characterizes the direction of dilution and the ordering of the overall odds ratio under several configurations.
Such findings have important implications for subgroup analysis and precision medicine.
Specifically, they explain why a treatment may appear effective within individual subgroups but diminish (or even disappear) when aggregated across the entire study population.
If ignored, this dilution effect may lead to misleading conclusions regarding treatment efficacy in RCTs.

Moreover, Theorem \ref{theorem-g-unchange} shows that the prevalence $\gamma^+$ does not affect the direction of dilution.
In other words, for given values of $p_2$, $p_4$, and $\lambda$, the way in which the two subgroups $g^+$ and $g^-$ are mixed does not alter the direction of the dilution effect.
The direction from which $OR_{\{g^+,g^-\}}$ approaches $\lambda$ depends instead on the configuration of the $g^-$ subgroup, determined by $p_2$ and $p_4$.
Specifically, if $p_4 > p_2$, then $OR_{\{g^+,g^-\}} < \lambda$; conversely, if $p_4 < p_2$, then $OR_{\{g^+,g^-\}} > \lambda$.

To develop an intuitive understanding of this theorem, we plot the overall odds ratio $OR_{\{g^+, g^-\}}$ as a function of the subgroup parameter.
Given the values of $p_2$, $p_4$, and $\gamma^+$, the overall odds ratio $OR_{\{g^+,g^-\}}$ can be expressed as a function of $p_1$, namely $OR_{\{g^+,g^-\}}=g(p_1)$, where
\begin{align}\label{eq-g}
g(x)&=\frac{f(\gamma^+f^{-1}(\lambda f(x))+(1-\gamma^+)p_4)}{f(\gamma^+x+(1-\gamma^+)p_2)} \nonumber \\
&=\frac{f(\gamma^+\frac{\lambda x}{(\lambda-1)x+1}+(1-\gamma^+)p_4)}{f(\gamma^+x+(1-\gamma^+)p_2)},\quad x\in(0,1).
\end{align}

\begin{figure}[htbp]
\centering
\subfigure[$p_2=0.1,p_4=0.25,\gamma^+=0.2$]{
\begin{minipage}[t]{0.5\linewidth}
\centering
\includegraphics[width=150pt]{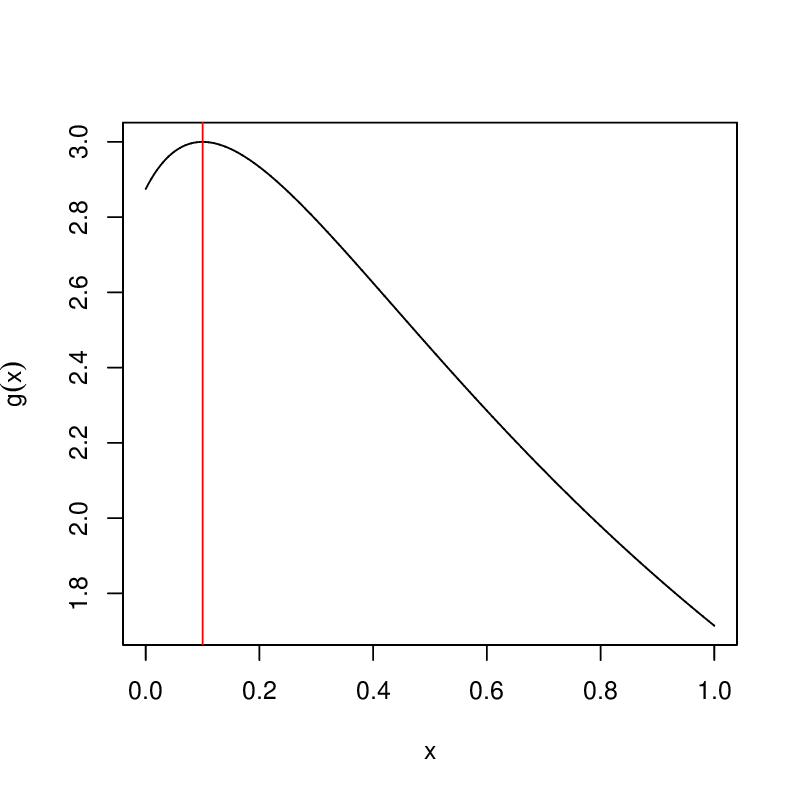}
\end{minipage}%
}%
\subfigure[$p_2=0.25,p_4=0.1,\gamma^+=0.2$]{
\begin{minipage}[t]{0.5\linewidth}
\centering
\includegraphics[width=150pt]{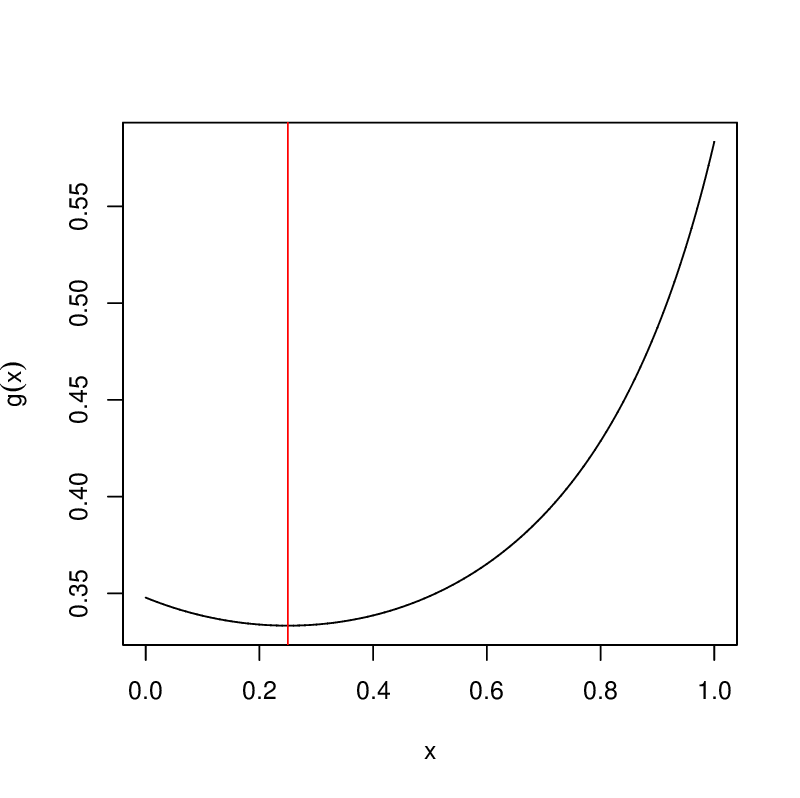}
\end{minipage}%
}%

\subfigure[$p_2=0.1,p_4=0.25,\gamma^+=0.5$]{
\begin{minipage}[t]{0.5\linewidth}
\centering
\includegraphics[width=150pt]{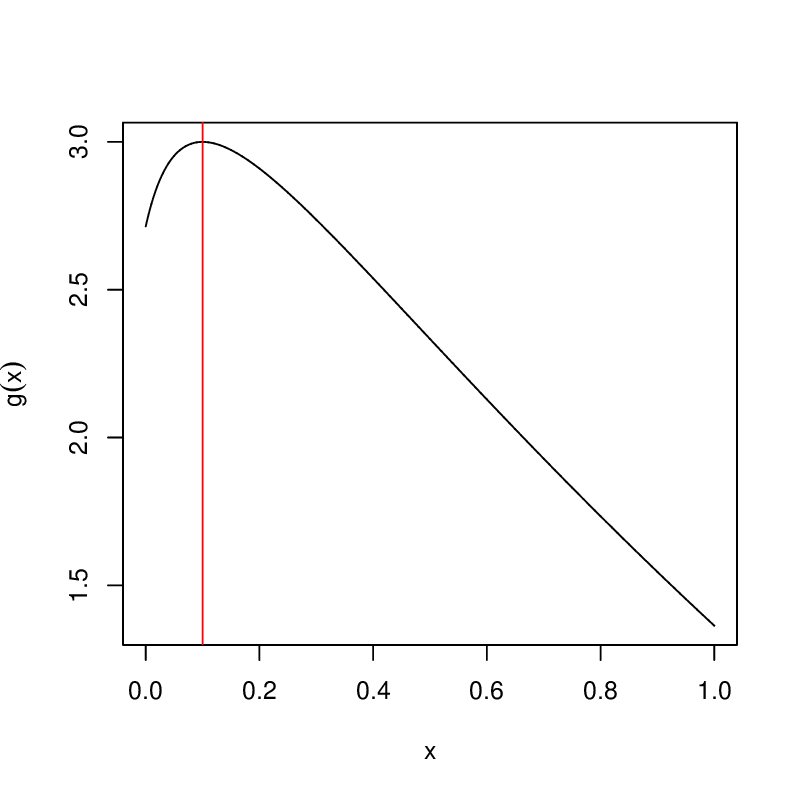}
\end{minipage}%
}%
\subfigure[$p_2=0.25,p_4=0.1,\gamma^+=0.5$]{
\begin{minipage}[t]{0.5\linewidth}
\centering
\includegraphics[width=150pt]{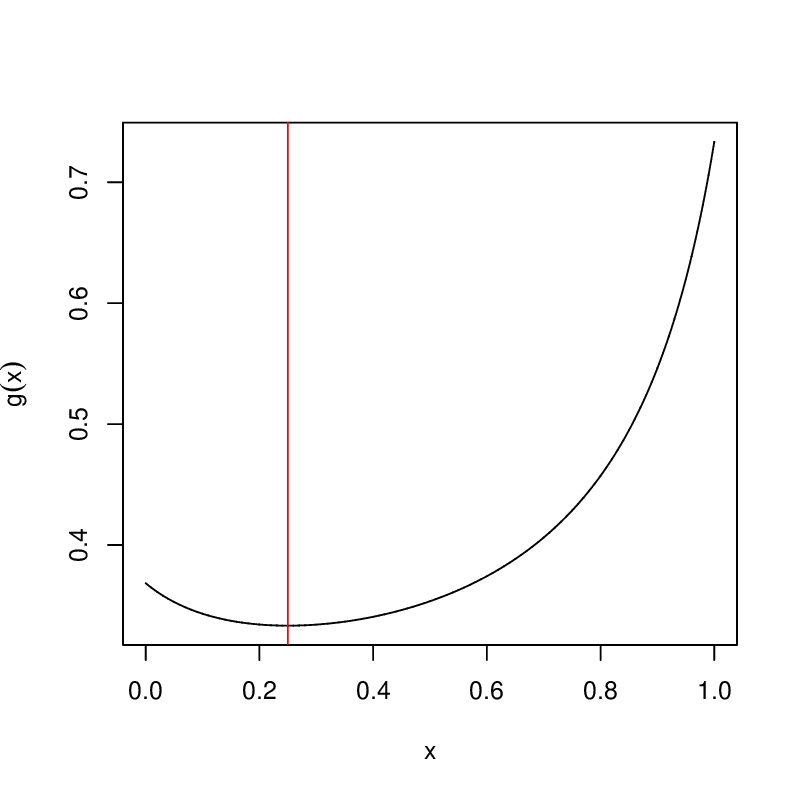}
\end{minipage}%
}%

\subfigure[$p_2=0.1,p_4=0.25,\gamma^+=0.8$]{
\begin{minipage}[t]{0.5\linewidth}
\centering
\includegraphics[width=150pt]{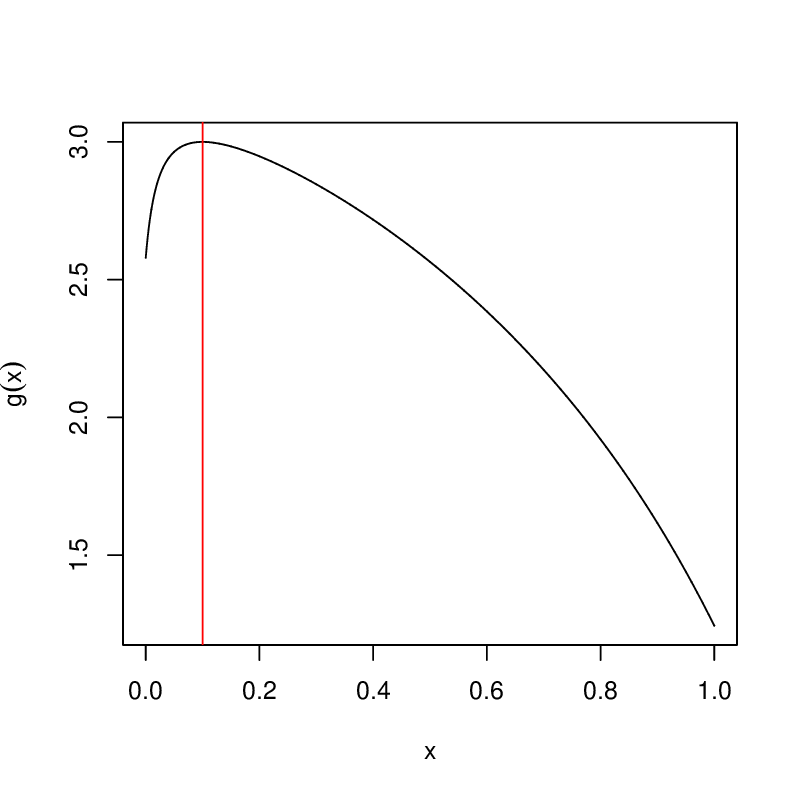}
\end{minipage}%
}%
\subfigure[$p_2=0.25,p_4=0.1,\gamma^+=0.8$]{
\begin{minipage}[t]{0.5\linewidth}
\centering
\includegraphics[width=150pt]{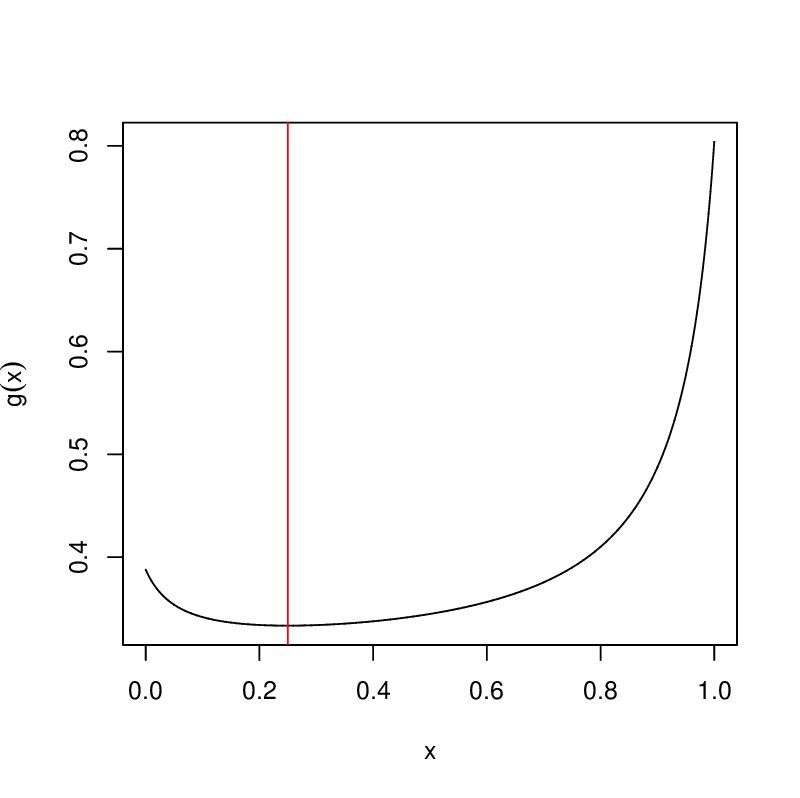}
\end{minipage}%
}%

\centering
\caption{The function $g(x)$ under different combination of $(p_2,p_4,\gamma^+)$}\label{figure-p-2-p-4-gamma}
\end{figure}

Figure \ref{figure-p-2-p-4-gamma} depicts the function $g(x)$, where $x$ denotes $p_1$ and the red line corresponds to the vertical line $x = p_2$, under different combinations of $(p_2, p_4, \gamma^+)$.
We observe that there exists exactly one extremum in the interval $(0,1)$.
Depending on the values of $p_2$ and $p_4$, this extremum may be either a maximum or a minimum.
As $\gamma^+$ increases, the shape of $g(x)$ exhibits slight changes.
Moreover, the figure provides clear visual confirmation of the result stated in Theorem \ref{theorem-g-unchange}.

From Figure \ref{figure-p-2-p-4-gamma}, if we draw a horizontal line $y=\beta$, this line may intersect with $g(x)$ at different points depending on the value of $\beta$.
Since $g(x)$ has exactly one extremum and is monotonic on either side of it, the number of intersections between $y=\beta$ and $g(x)$ is determined by $\beta$ and the values of $g(x)$ at the endpoints of its domain.
If the equation $g(x)=\beta$ has multiple solutions, then multiple configurations of $g^+$ satisfy the condition $OR_{g^+}=OR_{g^-}$.
The following theorem formalizes this result and provides criteria for determining whether such multiple solutions exist.
The detailed proof is provided in Appendix~S4 of the Supplementary Material.

\begin{theorem}\label{theorem-root}
Suppose that the $g^{-}$ subgroup is fixed and the $g^+$ subgroup  varies such that $OR_{g^+}^{(1)}=OR_{g^+}^{(2)}=OR_{g^-}=\lambda$. We have the following results:
\begin{align*}
\theta=\phi \ &\Leftrightarrow\ \left|p_4-\frac{1}{2}\right|=\left|p_2-\frac{1}{2}\right|,\\
\theta<\phi \ &\Leftrightarrow\ \left|p_4-\frac{1}{2}\right|>\left|p_2-\frac{1}{2}\right|,\\
\theta>\phi \ &\Leftrightarrow\ \left|p_4-\frac{1}{2}\right|<\left|p_2-\frac{1}{2}\right|,
\end{align*}
where
\begin{align*}
\theta=\lim_{p_1\rightarrow 0^+}g(p_1),\ \phi=\lim_{p_1\rightarrow 1^-}g(p_1).
\end{align*}

\end{theorem}

The practical implication of this theorem is that, given the overall $OR_{\{g^+,g^-\}}$ and the configuration of the $g^-$ subgroup, there may exist two distinct values of $p_1$ such that $OR_{g^+}=OR_{g^-}$ and, after mixing, lead to the same overall $OR_{\{g^+,g^-\}}$.
In other words, even if the $g^+$ subgroup shares the same odds ratio with $g^-$ subgroup and the overall odds ratio is fixed, the configuration of the $g^+$ subgroup cannot be uniquely identified.
Furthermore, by examining the relationship between $\theta$ and $\phi$, one can easily determine the number of intersection points between $y=\beta$ and $g(x)$, since $g(x)$ is monotonic on both intervals $(0,p_2)$ and $(p_2,1)$, where $p_2$ is the extremum of $g(x)$.
For example, when $p_2=0.1$, $p_4=0.25$, and $\gamma^+=0.2$: if $\beta\in(\theta,g(p_2))$, then $g(x)=\beta$ has two solutions; if $\beta\in(\phi,\theta]\cup {g(p_2)}$, it has exactly one solution; and for other values of $\beta$, no solution exists.

Even when we know that both $g^+$ and $g^-$ subgroups have the same odds ratio $\lambda$, and we also observe the overall odds ratio $OR_{\{g^+,g^-\}}=\beta$, this theorem shows that there may be two distinct values of $p_1$ (i.e., two different configurations of the treatment response in $g^+$) that result in the same overall odds ratio.
In practical terms, this means that the observed efficacy measure of the overall population does not uniquely determine the underlying subgroup behavior.
Attempts to infer subgroup-level behavior from the overall population may therefore be non-unique, leading to multiple possible subgroup configurations that are consistent with the same overall efficacy.
This result highlights a non-identifiability problem with important implications for subgroup analysis and personalized medicine.
In stratified populations (e.g., $g^+$ subgroup vs. $g^-$ subgroup), an overall odds ratio observed in the overall population does not necessarily correspond to a single underlying pattern of subgroup, even under the constraint $OR_{g^+}=OR_{g^-}$.
Without subgroup-specific information, the identification of the true subgroup may remain fundamentally ambiguous, even when data on the overall population and partial information on one subgroup are available.

\subsection{Approximate logic-respecting properties of the odds ratio and its connection to the relative response}

We first illustrate the relationship between the odds ratio and the relative response. It follows from (\ref{OR-condi}) that, for the $g^+$ subgroup, we have
\begin{align}\label{eq-30}
p_3=f^{-1}(OR_{g^+}\times f(p_1))=\frac{OR_{g^+}\times p_1}{(OR_{g^+}-1)p_1+1}.
\end{align}
Dividing both sides of (\ref{eq-30}) by $p_1$, then the relative response in the $g^+$ subgroup is
\begin{align*}
RR_{g^+}=\frac{p_3}{p_1}=\frac{OR_{g^+}}{(OR_{g^+}-1)p_1+1}.
\end{align*}
Similarly, we obtain
\begin{align*}
RR_{g^-}=\frac{p_4}{p_2}=\frac{OR_{g^-}}{(OR_{g^-}-1)p_2+1},\quad
RR_{\{g^+,g^-\}}=\frac{p_6}{p_5}=\frac{OR_{\{g^+,g^-\}}}{(OR_{\{g^+,g^-\}}-1)p_5+1}.
\end{align*}

In general, we have
\begin{align}\label{eq-33}
RR=\frac{OR}{(1-p)+p\times OR},
\end{align}
where $p$ denotes the conditional probability of response given the control ($C$) treatment. 
For a more detailed discussion of (\ref{eq-33}) in the setting of RCTs without considering subgroup, the reader is referred to \cite{ZK1998} and \cite{SK2008}.

Using (\ref{eq-33}), we can solve for the odds ratio as
\begin{align}\label{eq-34}
OR=\frac{RR\times(1-p)}{1-RR\times p}.
\end{align}
For the $g^+$ subgroup, the $g^-$ subgroup, and the combined population $\{g^+,g^-\}$, the corresponding odds ratios are given by
\begin{align}\label{eq-35}
OR_{g^+}&=\frac{RR_{g^+}\times(1-p_1)}{1-RR_{g^+}\times p_1},\quad OR_{g^-}=\frac{RR_{g^-}\times(1-p_2)}{1-RR_{g^-}\times p_2}, \nonumber \\
OR_{\{g^+,g^-\}}&=\frac{RR_{\{g^+,g^-\}}\times(1-p_5)}{1-RR_{\{g^+,g^-\}}\times p_5}.
\end{align}

\begin{figure}[htbp]
\centering
\subfigure[$(0,10)\times (0,10)$ region]{
\begin{minipage}[t]{0.5\linewidth}
\centering
\includegraphics[width=150pt]{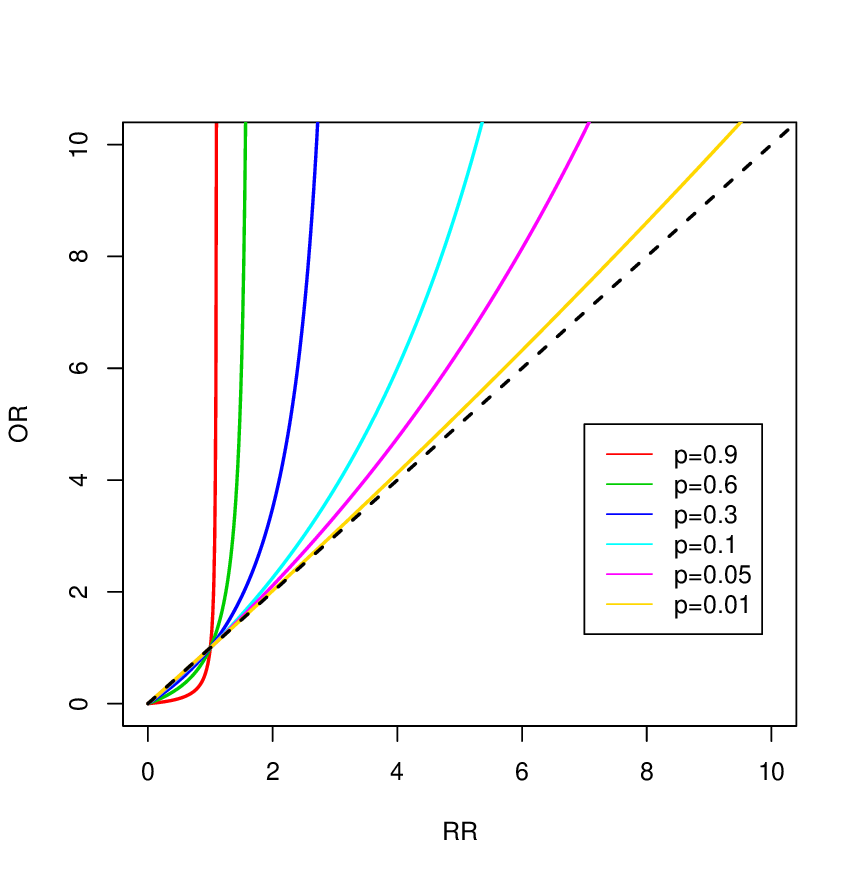}
\end{minipage}%
}%
\subfigure[$(0,2)\times (0,2)$ region]{
\begin{minipage}[t]{0.5\linewidth}
\centering
\includegraphics[width=150pt]{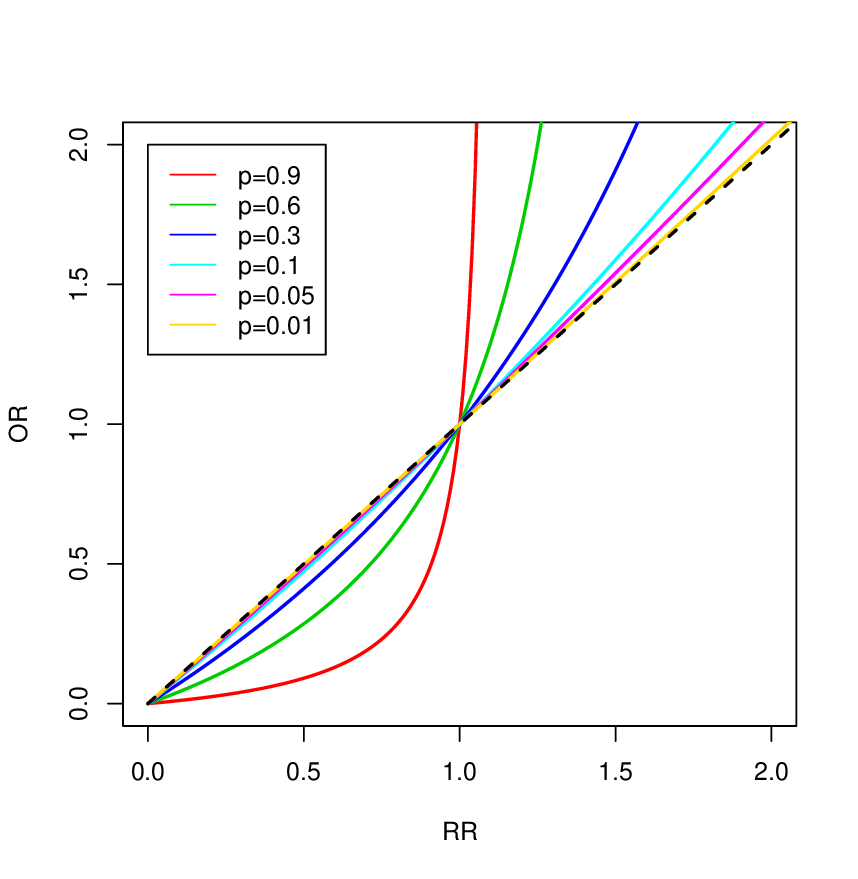}
\end{minipage}%
}%

\centering
\caption{The relationship between $RR$ and $OR$}\label{rr-or}
\end{figure}

For different values of $p\in(0,1)$, we plot the relationship between $OR$ and $RR$ according to (\ref{eq-34}).
In Figure \ref{rr-or}, panel (a) displays the $(0,10)\times(0,10)$ region, while panel (b) focuses on the $(0,2)\times(0,2)$ region.
The black dashed line is $y=x$.
We observe that the behavior of the function varies with $p$: curves grow more slowly when $0<RR<1$ and more rapidly when $RR>1$.
Notice that (\ref{eq-34}) always passes through the point $(1,1)$, regardless of the value of $p$.
As $p\rightarrow0^+$, the curve approaches the $y=x$ line.
Moreover, for a given $p$, $RR$ is defined on $(0,\frac{1}{p})$, with $x=\frac{1}{p}$ serving as a vertical asymptote.

\cite{PX2020} show that the odds ratio behaves as a logic-respecting measure when patients in the control ($C$) treatment are homogeneous across the $g^+$ and $g^-$ subgroups.
In other words, under certain conditions, the odds ratio appears to satisfy the logic-respecting property.
However, this phenomenon can be misleading, potentially creating an illusion of consistency.
It is therefore important to examine the possible causes of this illusion.
In this paper, we formally prove that the odds ratio for the combined population ${\{g^+,g^-\}}$ lies between the odds ratios of the individual subgroups $g^+$ and $g^-$ whenever the relative response is equal across the subgroups.
The detailed proof is provided in Appendix~S5 of the Supplementary Material.

\begin{theorem}\label{theorem-5}
If $RR_{g^+}=RR_{g^-}$, then $OR_{\{g^+,g^-\}}$ lies between $OR_{g^+}$ and $OR_{g^-}$.
\end{theorem}

This theorem indicates that conclusions drawn solely from the odds ratio may not be reliable.
Even when a new treatment shows apparent positive efficacy according to the odds ratio, the relative response may be homogeneous across subgroups.
In other words, the odds ratio can suggest a treatment benefit that is not reflected in the relative response, which directly compares outcome probabilities between groups.
This discrepancy underscores the importance of examining both measures when evaluating treatment effects, as relying exclusively on the odds ratio may lead to overly optimistic or misleading interpretations of efficacy.
Although the odds ratio is generally not a logic-respecting measure of efficacy, it can behave as if it were logic-respecting under certain conditions.
Theorem \ref{theorem-5} establishes a condition that differs from that in \cite{PX2020}, under which the odds ratio exhibits properties similar to those of a logic-respecting efficacy measure.

Next, we turn to the approximation of the odds ratio and examine how it can behave as an approximately logic-respecting efficacy measure.
Specifically, we show that under certain extreme conditions, the odds ratio can be regarded as an approximate representation for a logic-respecting efficacy measure.

Since the relative response is a logic-respecting efficacy measure, we have
\begin{align}\label{eq-39}
RR_{\{g^+,g^-\}}=\frac{p^C_{g^+}(R)}{p^C(R)}\times RR_{g^+} + \frac{p^C_{g^-}(R)}{p^C(R)}\times RR_{g^-}.
\end{align}
Substituting (\ref{eq-35}) into (\ref{eq-39}) yields
\begin{align}\label{eq-40}
&\frac{OR_{\{g^+,g^-\}}}{(1-p_5)+p_5\times OR_{\{g^+,g^-\}}} \nonumber \\
&=\frac{p^C_{g^+}(R)}{p^C(R)}\times \frac{OR_{g^+}}{(1-p_1)+p_1\times OR_{g^+}} + \frac{p^C_{g^-}(R)}{p^C(R)}\times \frac{OR_{g^-}}{(1-p_2)+p_2\times OR_{g^-}}.
\end{align}
If $p_1\rightarrow0^+,p_2\rightarrow0^+$, then $p_5\rightarrow0^+$ due to $p_5=\gamma^+p_1+(1-\gamma^+)p_2$. In this case, (\ref{eq-40}) can be approximated as
\begin{align*}
OR_{\{g^+,g^-\}}&\approx\frac{p^C_{g^+}(R)}{p^C(R)}\times OR_{g^+} + \frac{p^C_{g^-}(R)}{p^C(R)}\times OR_{g^-}.
\end{align*}
This expression can be rewritten as
\begin{align}\label{eq-60}
OR_{\{g^+,g^-\}}&\approx M^+\times OR_{g^+}+(1-M^+)\times OR_{g^-},
\end{align}
where $M^+=p^C_{g^+}(R)/p^C(R)$.
Thus, under these extreme conditions, $OR_{\{g^+,g^-\}}$ is approximately a weighted average of the subgroup-specific odds ratios $OR_{g^+}$ and $OR_{g^-}$.
The weight $M^+$ is not equal to the prevalence $\gamma^+$ in general.

While the relative response is always an exact weighted average of subgroup effects, the odds ratio can be interpreted in this way under rare-event conditions.
This finding demonstrates that the odds ratio behaves as approximately logic-respecting when event rates are small, thereby offering researchers a simple tool for sensitivity checks in practice.
The approximation in (\ref{eq-60}) is particularly relevant in settings where the event of interest is rare.
For instance, in clinical trials with low event rates (e.g., severe adverse events, or rare diseases), the values of $p_1, p_2$, and $p_5$ are small, and thus the overall odds ratio can be well approximated as a weighted average of the subgroup-specific odds ratios.
Similarly, in subgroup analyses where each subgroup exhibits low baseline risks, (\ref{eq-60}) provides an intuitive explanation of the relationship between subgroup and overall effects.
Moreover, this approximation helps to interpret situations where the relationship between subgroup-specific odds ratios and the overall odds ratio appears unintuitive in RCTs.
This may provide useful insights for subgroup analysis of rare diseases.

\section{An illustrative example}\label{sec-4}

In this section, we present a data-motivated illustrative example to show how the mechanisms described in Theorem \ref{theorem-g-unchange} and Theorem \ref{theorem-root} may arise in practice.
The dataset is drawn from REMAP-CAP, an international, adaptive, randomized clinical trial that evaluated the effects of tocilizumab and sarilumab on organ support-free days and survival in critically ill COVID-19 patients, conducted in the context of widespread concurrent corticosteroid use (see \cite{RemapCap2021}, \cite{REACT2021}, \cite{G2023}).
For clarity and convenience, we reorganize the data into Table \ref{T-ORR-2}, where $Rx$ denotes tocilizumab, and $C$ denotes usual care.
In this trial, patients were stratified into subgroups based on corticosteroid use at the time of randomization. 
The $g^-$ subgroup corresponds to patients without corticosteroid use, while the $g^+$ subgroup corresponds to patients with corticosteroid use.
In Table \ref{T-ORR-2}, the subgroup-specific odds ratios are close, but not exactly equal ($0.65$ versus $0.66$). To connect the empirical example with our theory, we construct a nearby illustrative configuration by setting the common subgroup odds ratio to $\lambda=0.65$ and keeping the $g^{-}$ subgroup fixed. This allows us to examine, in a data-motivated way, how the overall odds ratio changes as the configuration of the $g^{+}$ subgroup varies.

\begin{figure}
  \centering
  \includegraphics[width=250pt]{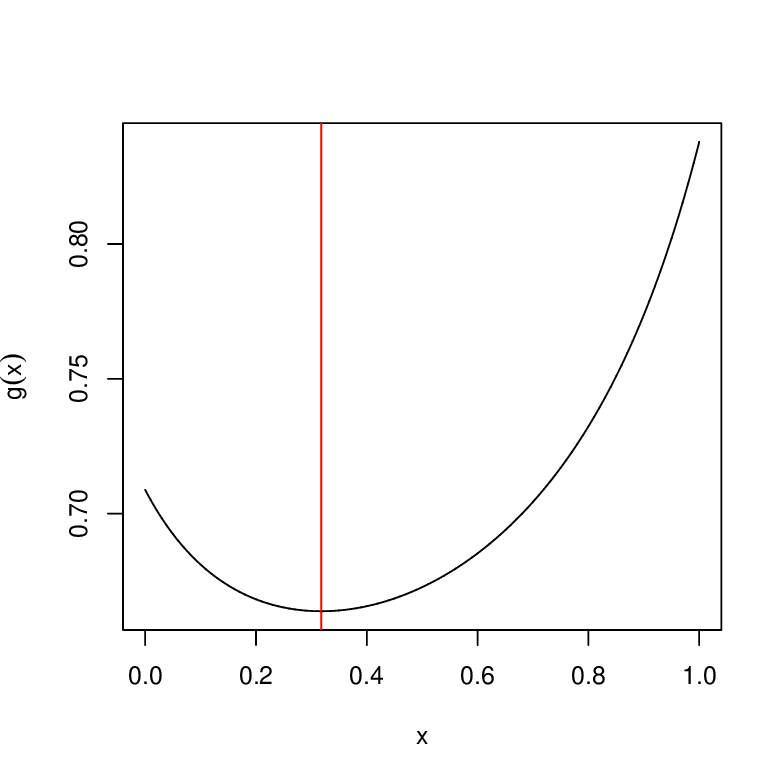}\\
  \caption{The relationship between $OR_{\{g^{+},g^{-}\}}$ and $p_1$ in a REMAP-CAP-motivated illustrative configuration with $\gamma^{+}=0.5$}
  \label{pic_real_data}
\end{figure}

The conditional probabilities used in this illustrative analysis are summarized in Table \ref{T-ORR-2-con}. Here we set the common subgroup odds ratio to $\lambda=0.65$, fix the $g^{-}$ subgroup at $(p_2,p_4)=(0.3178,0.2362)$, and vary the configuration of the $g^{+}$ subgroup as in Theorem \ref{theorem-g-unchange}. Because the prevalence of the $g^{+}$ subgroup in the target population is not available for this illustrative construction, we set $\gamma^{+}=0.5$ for presentation purposes. 
Accordingly, the curve of the overall odds ratio $OR_{\{g^+,g^-\}}$ is shown in Figure \ref{pic_real_data}, which is analogous to Figure \ref{figure-p-2-p-4-gamma}.
From this figure, we observe that when the overall odds ratio $OR_{\{g^+,g^-\}}$ lies within the interval $[0.7088,0.8378)\cup\{0.6638\}$, there exists a unique solution for the $g^+$ subgroup.
In contrast, when the overall odds ratio $OR_{\{g^+,g^-\}}$ falls within $(0.6638,0.7088)$, there are two distinct solutions for the $g^+$ subgroup.

This example illustrates how the theoretical mechanism can arise in a realistic parameter configuration motivated by trial data. Although the construction is illustrative, it helps connect the theoretical results to practical subgroup-analysis settings.

\begin{table}[htbp]
\centering
\footnotesize
\caption{REMAP-CAP clinical trial}
\label{T-ORR-2}
\setlength{\tabcolsep}{6pt}
\renewcommand{\arraystretch}{1.12}
\begin{tabular}{lcc@{\hspace{14pt}}cc}
\toprule
& \multicolumn{2}{c}{\(g^+\) subgroup (with corticosteroid)}
& \multicolumn{2}{c}{\(g^-\) subgroup (without corticosteroid)} \\
\cmidrule(lr){2-3} \cmidrule(lr){4-5}
& \(R\) & \(NR\) & \(R\) & \(NR\) \\
\midrule
\(Rx\) & 53 & 161 & 30 & 97 \\
\(C\)  & 73 & 144 & 41 & 88 \\
\midrule
Odds ratio & \multicolumn{2}{c}{0.65} & \multicolumn{2}{c}{0.66} \\
\bottomrule
\end{tabular}
\end{table}


\begin{table}[htbp]
\centering
\footnotesize
\caption{Conditional probabilities for the REMAP-CAP-motivated illustrative configuration}
\label{T-ORR-2-con}
\setlength{\tabcolsep}{6pt}
\renewcommand{\arraystretch}{1.12}
\begin{tabular}{lcc@{\hspace{14pt}}cc}
\toprule
& \multicolumn{2}{c}{\(g^+\) subgroup (with corticosteroid)}
& \multicolumn{2}{c}{\(g^-\) subgroup (without corticosteroid)} \\
\cmidrule(lr){2-3} \cmidrule(lr){4-5}
& \(R\) & \(NR\) & \(R\) & \(NR\) \\
\midrule
\(Rx\) & 0.2477 & 0.7523 & 0.2362 & 0.7638 \\
\(C\)  & 0.3364 & 0.6636 & 0.3178 & 0.6822 \\
\midrule
Odds ratio & \multicolumn{2}{c}{0.65} & \multicolumn{2}{c}{0.65} \\
\bottomrule
\end{tabular}
\end{table}


\section{Conclusions and outlook}\label{sec-5}

In this paper, we extended the work of \cite{LXDH2019} and \cite{Liu2020} by providing a deeper theoretical investigation into the behavior of the odds ratio and the relative response in subgroup analyses within the framework of RCTs.
We established several new theorems that characterize how the overall odds ratio changes when two subgroups are combined, with particular emphasis on its magnitude and direction.
We further demonstrated that, under a specific condition, the odds ratio can approximate a logic-respecting measure, thereby clarifying its relationship with the relative response.

Theorem \ref{theorem-g-unchange} and Theorem \ref{theorem-root} reveal that even if the odds ratio for the overall population and the configuration of one subgroup (e.g., $g^-$) are known, the configuration of the other subgroup (e.g., $g^+$) may still be non-unique.
That is, multiple values of $p_1$ may yield the same subgroup odds ratio $OR_{g^+}=OR_{g^-}$, yet lead to the same overall odds ratio $OR_{\{g^+,g^-\}}$.
In particular, it shows that different subgroup configurations may lead to the same observed overall effect. 
This non-uniqueness limits the ability to make reliable subgroup inferences solely based on the overall population.

In certain applications, it is possible to yield the same odds ratio but different relative response, reflecting different underlying event probabilities.
This discrepancy raises important questions regarding the choice of efficacy measure and the extent to which each measure is consistent with established medical reasoning.
The decision of whether odds ratio or relative response provides a more clinically logical interpretation has direct implications for the overall treatment evaluation.
The new findings in this paper further illustrate the limitations of the odds ratio as an efficacy measure in this subgroup setting, whereas the relative response provides a more transparent and generally more appropriate alternative, consistent with \cite{LXDH2019} and \cite{Liu2020}.

An important implication of our work is that it provides an interpretive framework for thinking about treatment efficacy when clinically meaningful subgroups may exist but are not directly identifiable in practice. In such settings, the observed overall odds ratio may conceal substantial subgroup heterogeneity, and the same overall effect may be compatible with different latent subgroup configurations. Therefore, when hidden treatment heterogeneity is plausible, the overall odds ratio alone should be interpreted with caution. More broadly, our results suggest that subgroup targeting should, whenever possible, be supported by subgroup-specific information or by sensitivity analyses that examine which latent subgroup configurations are compatible with the observed overall effect.

Furthermore, mathematical solutions may not correspond to realistic or clinically interpretable situations.
Careful consultation with clinicians is essential in these contexts to assess whether the observed inconsistency arises from issues in trial design, properties of the drug, or limitations of the chosen statistical framework.

\section*{Acknowledgment}

We are very grateful to Professor Jason C. Hsu for introducing us to the field of personalized medicine and for his important contributions to logic-respecting properties in subgroup analysis, which motivated this work.


\section*{Supplementary Material}
The proofs of the theorems presented in this paper are provided in Appendices~S.1--S.5 of the Supplementary Material.

\end{document}